\documentclass[proceedings]{rmaa}

\usepackage{rmaacite}

\title{Neutron Diffusion and Nucleosynthesis in an Inhomogeneous Big
           Bang Model}
\author{J. F. Lara
   \affil{Center for Relativity, UT, Austin}}

\fulladdresses{
   \item J. F. Lara:  University of Texas at Austin, Physics/Relativity, 
   Austin, TX 78712-1081 ( juan@einstein.ph.utexas.edu ).}

\shortauthor{Lara}
\shorttitle{Neutron Diffusion and Nucleosynthesis in IBBN}


\abstract{This article describes the production of primordial $ ^{4}$He nuclei
          in an inhomogeneous universe.  The baryon distribution is spherically
          symmetric and consists of a high density inner region and a low
          density outer region.  As the temperature decreases neutrons diffuse
          to the outer region until they are homogeneously distributed, and
          protons may be redistributed depending on how fast diffusion occurs. 
          Nucleosynthesis occurs earlier in the inner region and neutrons
          diffuse back to that region.  The rapidity of diffusion determines 
          how much $ ^{4}$He is ultimately produced.}

\listofauthors{J.~F.~Lara}
\indexauthor{Lara, J.~F.}

\begin{document}

\maketitle

\section{Introduction}
\label{sec:intro}

In 1999 the author wrote a big bang nucleosynthesis code corresponding to a
universe with an inhomogeneous baryon distribution.  The code envisions the 
universe as a lattice of spheres and treats one sphere as characteristic of the
rest of the universe.  The sphere has an initial radius ( the 
{\bf distance scale} $r_{i}$ ) of 25000 cm. at temperature $T = $ 100 GK.  The
sphere consists of a high density inner region and a low density outer region.
The {\bf boundary radius} $r_{b}$ between the regions is set to $r_{i}/2$ and
the {\bf density contrast} ratio $R_{\rho} = $ 800:1.  The baryon-to-photon 
ratio $\eta$ in this model is 3 $\times 10^{-10}$ ( $\eta_{10} = $ 3~).  The
sphere is divided into a core and 63 spherical shells.  

The code solve for the number density $n(i,s)$ of isotope species $i$ contained
in spherical shell $s$.  The evolution of $n(i,s)$ obeys the 
equation \cite{Kai98}

\begin{eqnarray*}
   \frac{dn(i,s)}{dt} & = & \frac{1}{n_{b}(s)} \sum_{j, k, l} N_{i} \left
     ( - \frac{n^{N_{i}}(i,s) n^{N_{j}}(j,s)}{N_{i}! N_{j}!} [ij] + 
           \frac{n^{N_{k}}(k,s) n^{N_{l}}(l,s)}{N_{k}! N_{l}!} [kl] \right ) \\
                      &   & - 3 \dot{\alpha} n(i,s) + \frac{1}{r^{2}}
    \frac{\partial}{\partial r} \left ( r^{2} D_{n} 
                        \frac{\partial n(i,s)}{\partial r} \right ) 
\end{eqnarray*}

\noindent The first term accounts for nuclear and weak reactions occuring 
between isotopes within shell $s$.  The second term corresponds to the 
expansion of the universe.  And the last term accounts for diffusion of isotope
$i$ between shells, though in the model only neutrons diffuse.

\begin{figure}
   \rotatebox{270}{\resizebox{3in}{6in}{\includegraphics{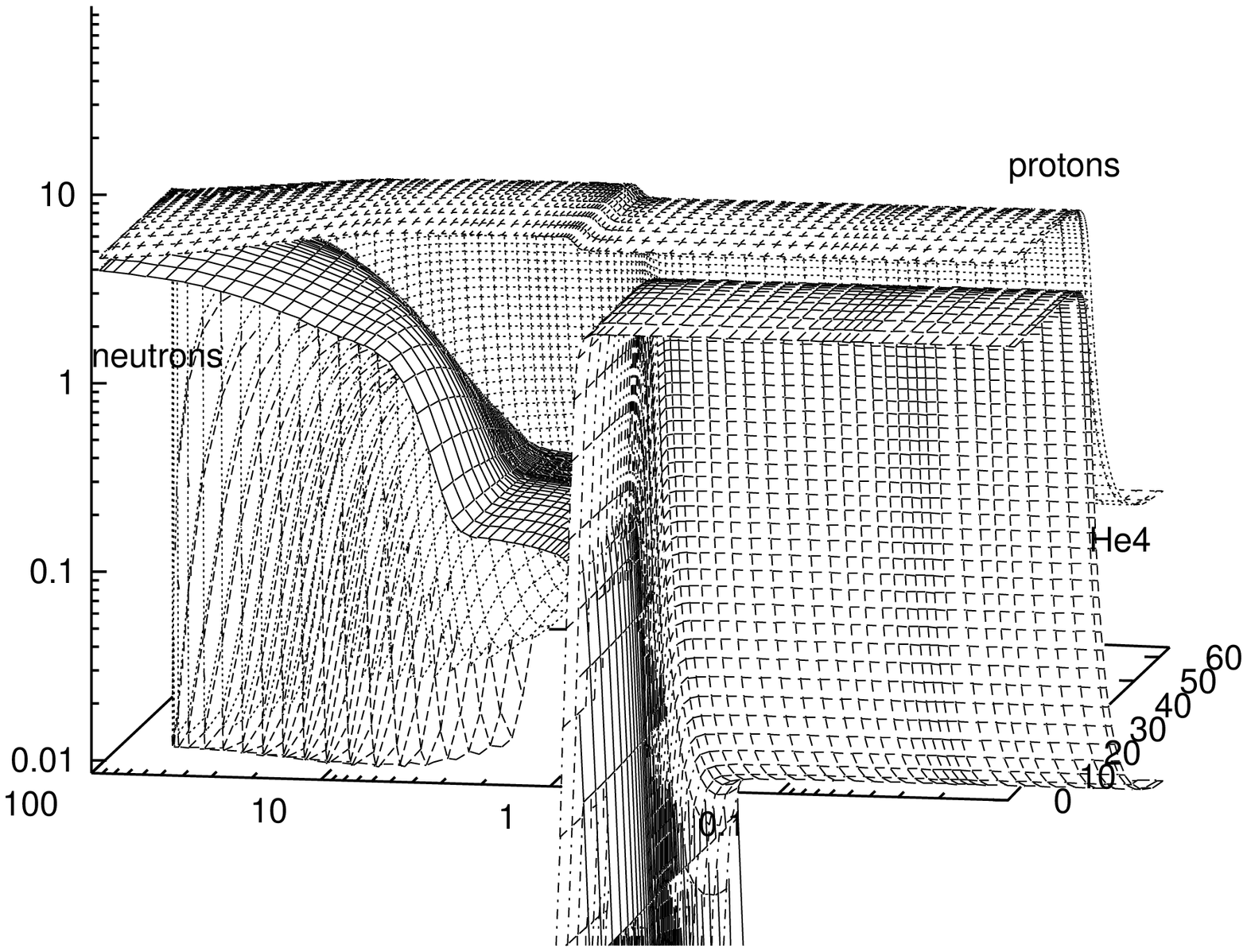}}}
   \caption{Number densities of neutrons, protons and $4 \times ^{4}$He}
   \leftline{\resizebox{3in}{3in}{\includegraphics{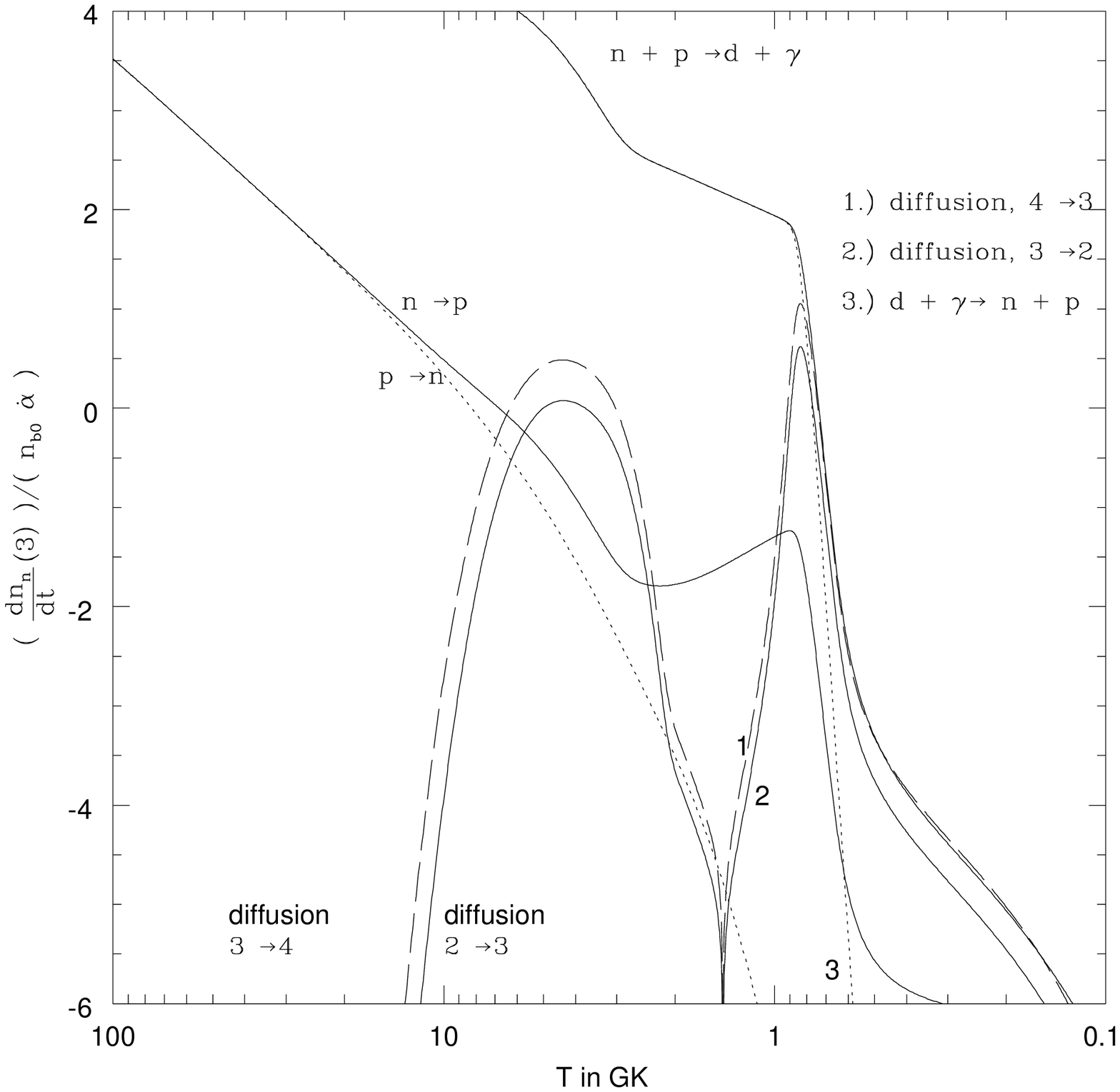}}
             \resizebox{3in}{3in}{\includegraphics{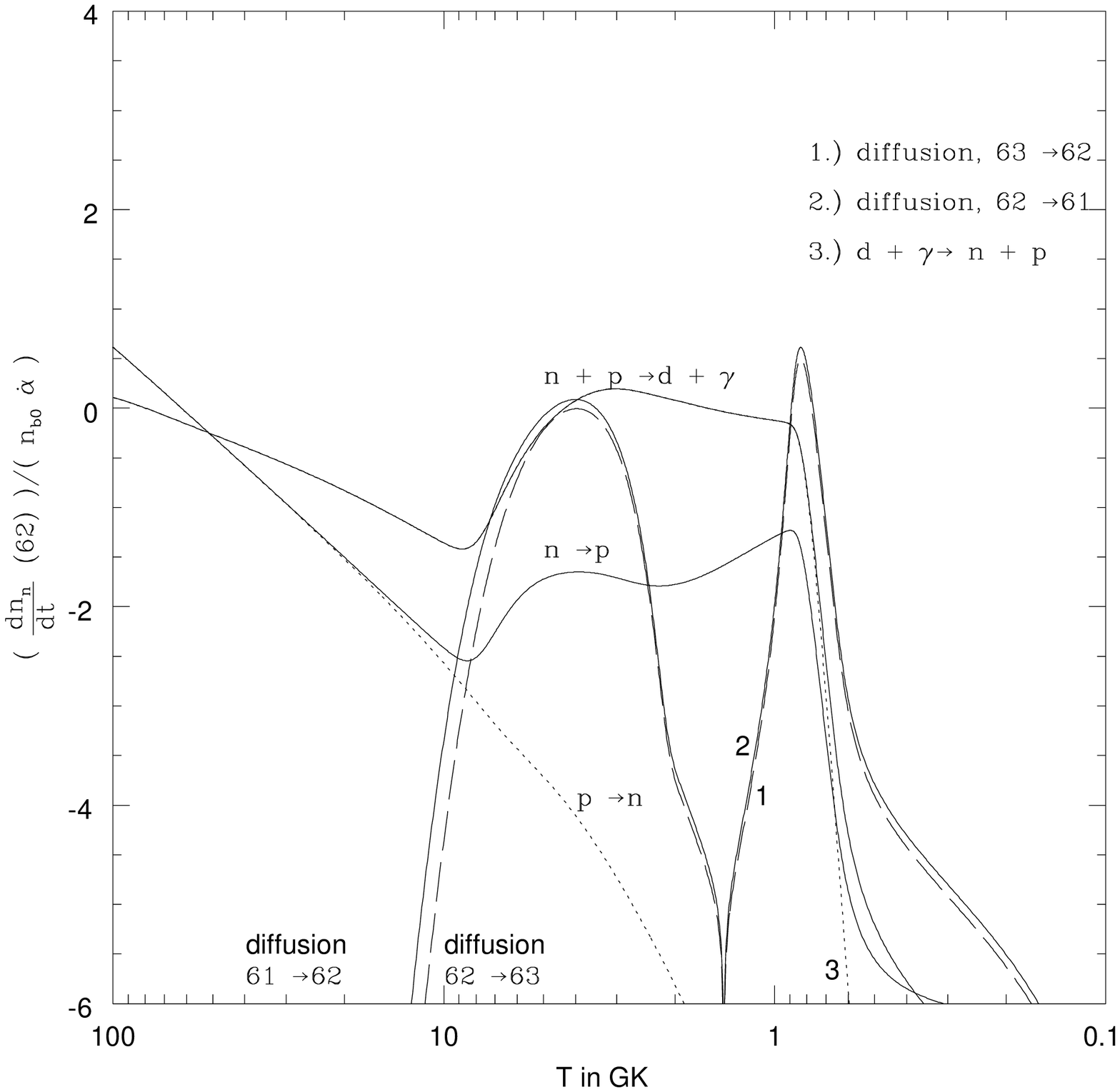}}}
   \caption{Neutron Reaction Rates ( shell 3 and shell 62 )}
\end{figure}

\section{Results for $r_{i} = $ 25000 cm, $\eta_{10} = $ 3}
\label{sec:Results}

Figure (1) shows the number densities of free neutrons ( $n({\mathrm n},s)$ ),
protons ( $n({\mathrm p},s)$ ) and 4 $\times ^{4}$He 
(~$4n( ^{4}{\mathrm He},s)$ ).  The horizontal axes correspond to the 
temperature $T$ from 100 GK down to 0.01 GK and the distance scale $r_{i}$ 
normalized to 64 dimensionless units, and the vertical axis is number density 
$n(i,s)$ normalized to the initial baryon density ( $n_{b0}$ ).  At $T = $ 100 
GK number densities in the inner shells vs. the outer shells are large.  Weak
reactions convert neutrons to protons until they no longer become significant
at around $T = $ 13 GK.  Figure (2a) shows the neutron-proton conversion rates 
for shell 3, a high density inner shell ( normalized to $n_{b0}$ and to the
expansion rate of the universe $\dot{\alpha}$ ).  Figure (2a) also shows that
around $T = $ 6 GK neutrons start to diffuse significantly into the lower
density outer shells.  But since the conversion rates are no longer effective
at this temperature the protons in the inner shells remain there.  At around
$T = $ 2.5 GK the neutrons are homogeneously distributed throughout the whole
sphere.

Nucleosynthesis occurs at about $T = $ 0.9 GK, when the nuclear reaction 
n + p $\leftrightarrow$ d + $\gamma$ falls out of equilibrium.  Other nuclear
reactions then build heavier nuclei, but most baryons wind up as part of 
$ ^{4}$He nuclei.  Because of higher density nucleosynthesis occurs earlier in
the inner shells, depleting neutrons.  So neutrons diffuse back into the inner
regions as shown on the right of Figure (2a).  Figure (2b) shows the same 
reaction rates for shell 62, a low density outer shell.  The rate for n + p
$\leftrightarrow$ d + $\gamma$ is lower and flatter than in Figure (2a).  
Figure (1) shows that most neutrons back diffuse into the inner shells and then
undergo nucleosynthesis, leading to $ ^{4}$He production overwhelmingly 
concentrated in the inner shells.  

The final mass fraction $X_{ ^{4}{\mathrm He}}$ of the whole model is 0.243664.
This is the result for $r_{i} = $ 25000 cm and $\eta_{10} = 3$  Table (1) shown
below lists $X_{ ^{4}{\mathrm He}}$ for $r_{i} = $ 100 cm up to 6.3 
$\times 10^{6}$ cm.  For $r_{i}$ less than 25000 cm neutron diffusion occurs 
early enough to coincide at least partly with the time when the conversion
reactions are significant.  In the inner regions the reactions convert protons
into neutrons that then diffuse to the outer regions where they get converted 
back to protons.  But as $r_{i}$ gets larger more protons remain in the inner
regions and $X_{ ^{4}{\mathrm He}}$ increases.  $r_{i} = $ 25000 cm is the
turning point when all the protons in the innermost regions remain there.  For
larger $r_{i}$ the back diffusing neutrons can't reach the innermost regions
before nucleosynthesis is completed.  So with fewer neutrons in the inner
regions $X_{ ^{4}{\mathrm He}}$ decreases.  At $r_{i} = 7.9 \times 10^{5}$ cm
though diffusion becomes significant at $T = $ 0.9 GK, right during 
nucleosynthesis.  For larger $r_{i}$ nucleosynthesis occurs before neutron 
diffusion and the inner regions, with high proton and neutron densities, 
produce enough $ ^{4}$He to raise $X_{ ^{4}{\mathrm He}}$ significantly.

\begin{center}

\begin{tabular}{r|c}
      $r_{i}$ in cm.     & $X_{ ^{4}{\mathrm He}}$ \\ \hline
        $10^{2}$         &       0.239029657 \\ 
 1.58489 $\times 10^{3}$ &       0.239369667 \\
 2.51189 $\times 10^{4}$ &       0.243663794 \\
        $10^{5}$         &       0.241847338 \\
 7.94328 $\times 10^{5}$ &       0.229781933 \\
 6.30957 $\times 10^{6}$ &       0.250711334 \\
\end{tabular}

\vspace{0.2in}

   {\bf Table (1)}:  $ ^{4}$He, where $\eta = 3.0 \times 10^{-10}$.
\end{center}

\section{Conclusions and Future Research}
\label{sec:Conclusions}

Figure (3a) shows $X_{ ^{4}{\mathrm He}}$ for a range of $\eta$ from $10^{-10}$
to $3 \times 10^{-9}$ as well as the range of $r_{i}$ listed in Table (1)  For
all values of $\eta$ shown one can see the turning points at which 
$X_{ ^{4}{\mathrm He}}$ increases and then decreases and then increases again.
The code can generate contour maps for 67 other isotope species as well.  
Comparing the contour maps of $X_{ ^{4}{\mathrm He}}$, 
$Y({\mathrm d})/Y({\mathrm p})$ ( $Y({\mathrm d})$ is the abundance of 
deuterium ), and $Y( ^{7}{\mathrm Li} + ^{7}{\mathrm Be})/Y({\mathrm p})$ to 
the $2\sigma$ ranges of recent observations ( IT, 1998; \cite{Izo98} BT, 1998;
\cite{Bur98} PWSN, 1998 \cite{Pin98} ).

\begin{eqnarray*}
                          X_{ ^{4}{\mathrm He}} & = & 0.244 \pm 0.002 \\
          Y({\mathrm d})/Y({\mathrm d}) & = & ( 3.4 \pm 0.3 ) \times 10^{-5} \\
   \log [ Y( ^{7}{\mathrm Li})/Y({\mathrm d}) ] & = & -9.45 \pm 0.20
\end{eqnarray*}
   
\noindent the author determined an observational range for $\eta$ from 4.7
$\times 10^{-10}$ to 6 $\times 10^{-10}$.  

A planned article will describe $ ^{4}$He production in greater detail, 
describing results for values of $r_{i}$ other than 25000 cm.  $R_{\rho}$, 
$r_{b}$ and the geometry of the model can be varied in the code.  In the future
the author will look at results with different values for these parameters.  
Also, the code has written into it subroutines to calculate the neutrino 
heating effect.  Electrons and positrons annihilate each other from $T = $ 5 GK
to $T = $ 1 GK.  Through annihilation and scattering they can transfer some 
energy to neutrinos, which in turn can alter the neutron-proton conversion 
reaction rates and ultimately $ ^{4}$He production.  Figure (3b) shows the 
change $\Delta X_{ ^{4}{\mathrm He}}$ of the mass fraction of $ ^{4}$He for 
the same ranges of $\eta$ and $r_{i}$ as in Figure (3a).  
$\Delta X_{ ^{4}{\mathrm He}}$ imitates $X_{ ^{4}{\mathrm He}}$, but not 
exactly.  The change also remains within an order of $10^{-4}$.  The author 
will determine if the change is equivalent to a constant shift of Figure (3a)
to the left, and if the coding for the neutrino heating effect is correctly
represented in the code.

\begin{figure}
   \leftline{\rotatebox{270}{\resizebox{3in}{3in}
                             {\includegraphics{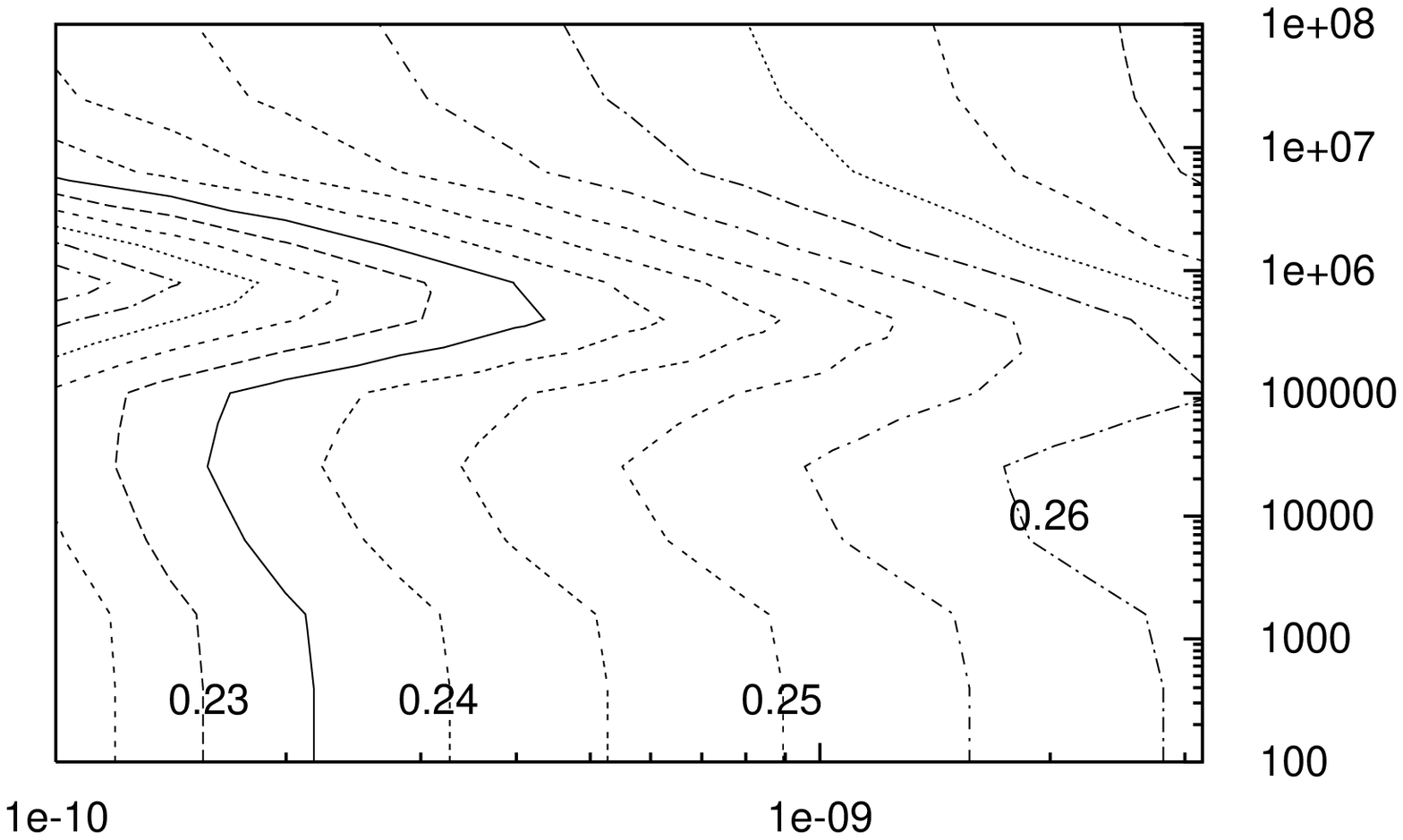}}}
             \rotatebox{270}{\resizebox{3in}{3in}
                             {\includegraphics{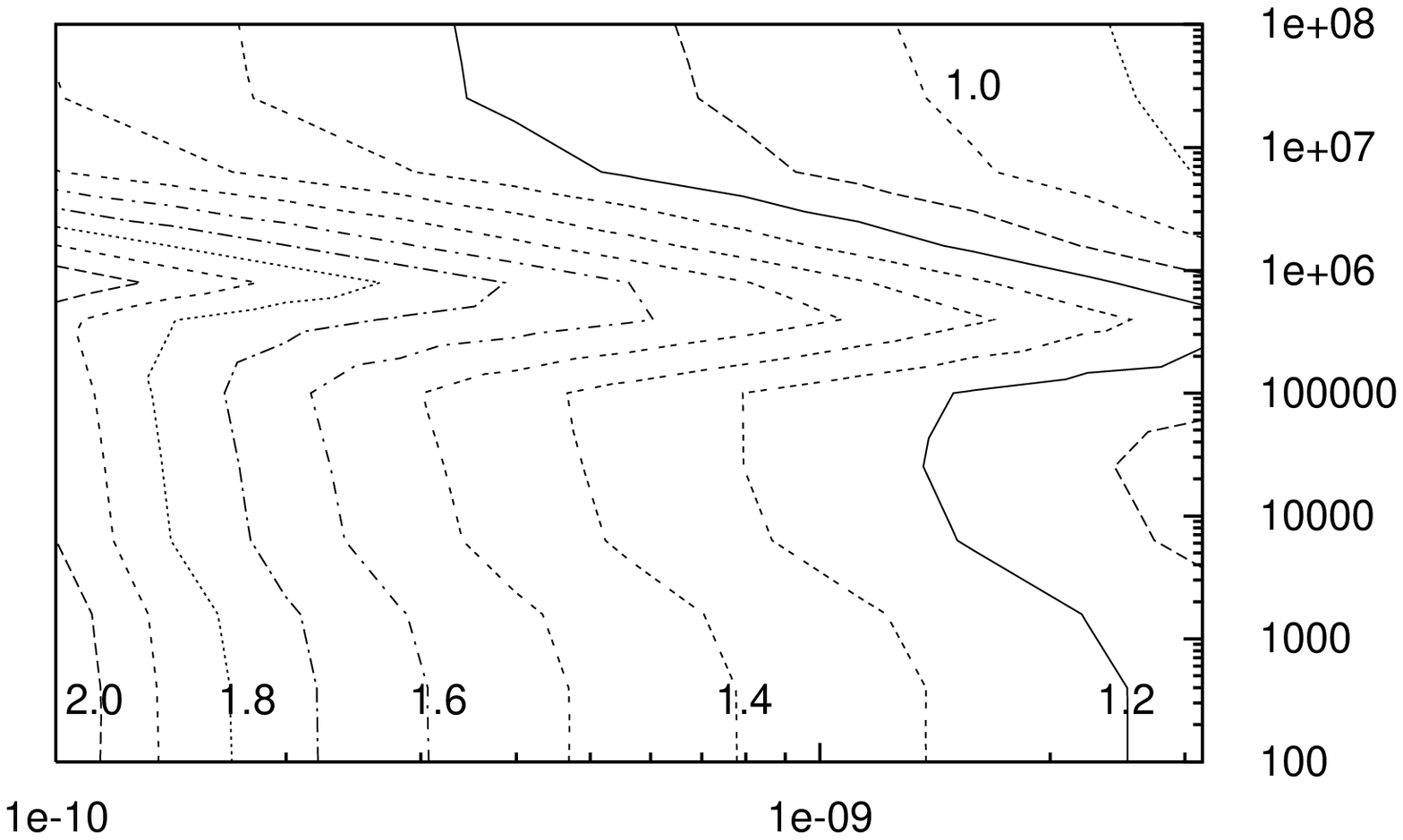}}}}
   \caption{$X_{ ^{4}{\mathrm He}}$ and $10^{4} \times \Delta 
            X_{ ^{4}{\mathrm He}}$}
\end{figure}


\begin{thebibliography}

\bibitem[Burles \& Tytler<1998>]{Bur98}
   Burles, S. \& Tytler, D.  astro-ph/9803071

\bibitem[Izotov \& Thuan<1998>]{Izo98}
   Izotov, Y.~I. \& Thuan, T.~X.  1998, ApJ {497}, 227

\bibitem[Kainulainen et al.{}<1998>]{Kai98}
   Kainulainen, K., Devine, J. \& Sihvola, E.  astro-ph/9807098

\bibitem[Pinsonneault et at.{}<1998>]{Pin98}
   Pinsonneault, M.~H., Walker, T.~P., Steigman, G. \& Narayanan, V.~K.
   astro-ph/9802315

\end{thebibliography}
\end{document}